\documentclass{article}

\usepackage[preprint]{neurips_2024}

\usepackage[utf8]{inputenc} % allow utf-8 input
\usepackage[T1]{fontenc}    % use 8-bit T1 fonts
\usepackage{hyperref}       % hyperlinks
\usepackage{url}            % simple URL typesetting
\usepackage{booktabs}       % professional-quality tables
\usepackage{amsfonts}       % blackboard math symbols
\usepackage{nicefrac}       % compact symbols for 1/2, etc.
\usepackage{microtype}      % microtypography
\usepackage{xcolor}         % colors
\usepackage{diagbox}

\title{Measurement Embedded Schrödinger Bridge for Inverse Problems}

\author{%
  Yuang Wang \\
  Department of Engineering Physics, Tsinghua University, Beijing, China \\
  Department of Radiology, Massachusetts General Hospital and Harvard Medical School, Boston, USA
  \AND
  Pengfei Jin\\
    Department of Radiology, Massachusetts General Hospital and Harvard Medical School, Boston, USA \\
  \AND
  Siyeop Yoon\\
    Department of Radiology, Massachusetts General Hospital and Harvard Medical School, Boston, USA \\
  \AND
  Matthew Tivnan\\
    Department of Radiology, Massachusetts General Hospital and Harvard Medical School, Boston, USA \\ 
  \AND
  Quanzheng Li\\
    Department of Radiology, Massachusetts General Hospital and Harvard Medical School, Boston, USA \\
  \AND
  Li Zhang\\
    Department of Engineering Physics, Tsinghua University, Beijing, China \\
  \AND
  Dufan Wu\thanks{Corresponding author: dwu6@mgh.harvard.edu}\\
    Department of Radiology, Massachusetts General Hospital and Harvard Medical School, Boston, USA \\
}
\usepackage{amsmath}
\newtheorem{theorem}{Theorem}
\usepackage{algorithm}
\usepackage{algorithmic}
\usepackage{graphicx}
\usepackage{verbatim} 
\setcitestyle{numbers,square}
\begin{document}

\maketitle

\begin{abstract}
Score-based diffusion models are frequently employed as structural priors in inverse problems. However, their iterative denoising process, initiated from Gaussian noise, often results in slow inference speeds. The Image-to-Image Schrödinger Bridge (I$^2$SB), which begins with the corrupted image, presents a promising alternative as a prior for addressing inverse problems. In this work, we introduce the Measurement Embedded Schrödinger Bridge (MESB). MESB establishes Schrödinger Bridges between the distribution of corrupted images and the distribution of clean images given observed measurements. Based on optimal transport theory, we derive the forward and backward processes of MESB. Through validation on diverse inverse problems, our proposed approach exhibits superior performance compared to existing Schrödinger Bridge-based inverse problems solvers in both visual quality and quantitative metrics.
\end{abstract}

\section{Introduction}

Inverse problems are prevalent across various scientific and engineering disciplines with the aim of uncovering an unknown signal from observed measurements. The inherent complexity arises from their ill-posed nature, where multiple solutions can plausibly explain the observed data. To address this challenge, score-based diffusion models \cite{song2019generative,ho2020denoising,song2020score} are commonly employed as a structural prior, facilitating the derivation of meaningful solutions. Numerous techniques have been developed to integrate data consistency into the sampling process of diffusion models and maintain the sample path on the data manifold\cite{chung2023fast,chung2022diffusion,chung2022improving,song2023solving}. While diffusion-based inverse problem solvers have achieved success in image restoration tasks, their inference speed is often hampered by the need for a large number of iterative denoising steps to generate clean images from pure Gaussian noise.

Instead of starting from Gaussian noise, Schrödinger Bridges\cite{kim2023unpaired,delbracio2023inversion,chen2021likelihood,chen2021stochastic,de2021diffusion} establish diffusion bridges between the distributions of clean and corrupted images. By initializing the diffusion process with the corrupted image, which is closer to the clean one compared to Gaussian noise, Schrödinger Bridges offer a promising approach for generating high-quality conditional samples with fewer diffusion steps. One notable example is the Image-to-Image Schrödinger Bridge (I$^2$SB)\cite{liu20232}, which models the diffusion bridge between paired clean and corrupted samples. This model enables efficient training by connecting it with the standard score-based diffusion model. Recently, the data Consistent Direct Diffusion Bridge (CDDB)\cite{chung2024direct} has further enhanced the performance of I$^2$SB in both perceptual ability and fidelity to ground truth by incorporating data consistency techniques similar to those used in diffusion models.

In our research, we propose the Measurement Embedded Schrödinger Bridge (MESB) for solving inverse problems. Unlike CDDB that constructs Schrödinger Bridges between clean and corrupted image distributions and incorporates data consistency during sampling, MESB directly establishes Schrödinger Bridges between the distribution of corrupted images and the distribution of clean images given observed measurements. Based on optimal transport theory, we derive the forward and backward processes of MESB, inherently incorporating data consistency into the backward process. Our method is validated across diverse inverse problems, demonstrating superior performance compared to existing Schrödinger Bridge-based inverse problem solvers.

\section{Preliminaries}
Notation: Consider a d-dimensional stochastic process denoted by $X_t \in \mathbb{R}^d$, where $t \in [0,1]$ indexes the process. We use $X_{clean}$ to represent a sample from the clean image distribution $q_{clean}$ and $X_{corrupt}$ for a sample from the corrupted image distribution $q_{corrupt}$. Let $N$ be the number of generative steps. We define discrete generative time steps as $0 = t_0 < t_1 < \dots < t_n < \dots < t_N = 1$, and use the shorthand $X_n \equiv X_{t_n}$. 
\subsection{Schrödinger Bridge}
The Schrödinger Bridge is an entropy-regularized optimal transport approach\cite{leonard2013survey} that constructs diffusion bridges between two arbitrary distributions $p_A$ and $p_B$. It is defined by the following forward and backward stochastic differential equations (SDEs):
\begin{equation}
dX_t=\left[f_t+\beta_t\nabla_{X_t}\log\Psi\left(X_t,t\right)\right]dt+\sqrt{\beta_t}dw_t,
\label{eq:sde_forw}
\end{equation}
\begin{equation}
dX_t=\left[f_t-\beta_t\nabla_{X_t}\log\hat{\Psi}\left(X_t,t\right)\right]dt+\sqrt{\beta_t}d\overline{w}_t.
\label{eq:sde_back}
\end{equation}
In these equations, $X_0$ is sampled from $p_A$, $X_1$ is sampled from $p_B$, $f_t$ represents the basic drift of $X_t$, $\beta_t$ determines the diffusion speed, and $w_t$ and $\overline{w}_t$ are the Wiener process and its reversed counterpart. To ensure that the path measure induced by the forward SDE (\ref{eq:sde_forw}) is almost surely equal to the one induced by the reverse SDE (\ref{eq:sde_back}), the time-varying energy potentials $\Psi$ and $\hat{\Psi}$ must satisfy the following coupled partial differential equations (PDEs):
\begin{equation}
\left\{\begin{array}{l} \frac{\partial\Psi}{\partial t}=-\nabla\Psi^\mathsf{T}f-\frac{1}{2}\beta\triangle\Psi,\\
\frac{\partial\hat{\Psi}}{\partial t}=-\nabla\cdot\left(\hat{\Psi} f\right)+\frac{1}{2}\beta\triangle\hat{\Psi},
\end{array}\right.
\label{eq:PDE}
\end{equation}
subject to the margin conditions:
\begin{equation}
\Psi\left(X_0,0\right)\hat{\Psi}\left(X_0,0\right)=p_A\left(X_0\right),\Psi\left(X_1,1\right)\hat{\Psi}\left(X_1,1\right)=p_B\left(X_1\right).
\label{eq:PDE_margin}
\end{equation}
\subsection{Score-based Diffusion Models}
Score-based diffusion models\cite{ho2020denoising}, as a special case of the Schrödinger Bridge, specify $p_A$ as $q_{clean}$ and $p_B$ as the Gaussian distribution $\mathcal{N}(0, I)$, using a linear $f_t$ with respect to $X_t$. In this setup, $\Psi$ simplifies to $1$, and $\nabla \log \hat{\Psi}$ becomes the score function and can be learned through denoising score matching\cite{vincent2011connection}.
\subsection{Image-to-Image Schrödinger Bridge}
The coupling of $\Psi$ and $\hat{\Psi}$ in the margin conditions (\ref{eq:PDE_margin}) can lead to computational challenges when setting $p_A$ as $q_{clean}$ and $p_B$ as $q_{corrupt}$. I$^2$SB\cite{liu20232} offers an approach to decouple $\Psi$ and $\hat{\Psi}$. Instead of treating the clean image distribution as a continuous manifold, it is represented as a sum of delta functions centered at each clean image. By setting $p_A(X_0)$ as $\delta(X_0 - X_{clean})$, $\Psi$ and $\hat{\Psi}$ become decoupled, enabling the establishment of a Schrödinger Bridge for each clean and corrupted image pair. With $f$ set to $0$, I$^2$SB can be trained as efficiently as score-based diffusion models. 

Specifically, in the forward process, $X_t$ is sampled from the distribution $q\left(X_t|X_0,X_1\right)$:
\begin{equation}
q\left( X_t|X_0,X_1\right)=\mathcal N\left(X_t;\frac{\overline{\sigma}_t^2}{\overline{\sigma}_t^2+\sigma_t^2}X_0+\frac{{\sigma}_t^2}{\overline{\sigma}_t^2+\sigma_t^2}X_1,\frac{{\sigma}_t^2\overline{\sigma}_t^2}{\overline{\sigma}_t^2+\sigma_t^2}I\right),
\label{eq:mu_t}
\end{equation}
where ${\sigma}_t^2=\int_0^t\beta_\tau d\tau$ and $\overline{\sigma}_t^2=\int_t^1\beta_\tau d\tau$ represent variances accumulated from either side. The network $\epsilon_\theta$ can be efficiently trained to predict the difference between $X_t$ and $X_0$ by minimizing the loss function:
\begin{equation}
\theta^*=\arg\min_{\theta}E_{X_0, X_1}E_{t\sim \mathcal{U}[0,1],X_t\sim q\left(X_t|X_0,X_1\right)}\Vert \epsilon_{\theta}\left(X_t,t\right)-\frac{X_t-X_0}{\sigma_t} \Vert.
\label{eq:loss}
\end{equation}
In the reverse process, I$^2$SB sets $X_N$ as the corrupted image $X_{corrupt}$ and iteratively approaches $X_0$. In the step from $X_n$ to $X_{n-1}$, $\hat{X}_0$, the expected mean of $X_0$, is first calculated using the trained network $\epsilon_{\theta^*}$ and $X_n$:
\begin{equation}
\hat{X}_0 = X_{n}-\sigma_{n}\epsilon_{\theta^*}\left(X_{n},t_{n},y\right),
\label{eq:x0_hat}
\end{equation}
where $\sigma_n \equiv \sigma_{t_n}$. Subsequently, $X_{n-1}$ is sampled from the DDPM posterior $p\left(X_{n-1}|\hat{X}_0, X_n\right)$, expressed as:
\begin{equation}
p\left(X_{n-1}|X_0, X_n\right)=\mathcal{N}\left(X_{n-1};\frac{\alpha_{n-1}^2}{\alpha_{n-1}^2+\sigma_{n-1}^2}X_0+\frac{\sigma_{n-1}^2}{\alpha_{n-1}^2+\sigma_{n-1}^2}X_n,\frac{\sigma_{n-1}^2\alpha_{n-1}^2}{\alpha_{n-1}^2+\sigma_{n-1}^2}I\right),
\label{eq:p}
\end{equation}
where $\alpha_{n-1}^2=\int_{t_{n-1}}^{t_n}\beta_\tau d\tau$ denotes the accumulated variance between consecutive time steps $t_{n-1}$ and $t_n$.

%Conclude: SB is a method ... pa  pb. However, \{motivation for invers promblem AX=y\},下一个section，我们提出模型，y的先验引入pa。
\section{Method}
In this section, we propose MESB for inverse problems, establishing Schrödinger Bridges between the distribution of corrupted images and the distribution of clean images given measurements. We embed measurements into the margin conditions \ref{eq:PDE_margin}, obtain $\Psi$ and $\hat{\Psi}$ by solving the PDEs (\ref{eq:PDE}) and (\ref{eq:PDE_margin}), and derive the forward and backward processes of MESB using SDEs (\ref{eq:sde_forw}) and (\ref{eq:sde_back}). All proofs are provided in Appendix A.1.
\subsection{Measurement Embedded Schrödinger Bridge for Inverse Problems}
In the context of inverse problems, we start with $X_{corrupt}$ and seek to recover $X_{clean}$ based on the measurement $y$, which is related through the system matrix $A$ and detector noise $n$, as described by the equation:
\begin{equation}
y=AX_{clean}+n.
\label{eq:inverse problem}
\end{equation}
To address this challenge, we propose MESB to establish Schrödinger Bridges that map each corrupted image to its corresponding clean image distribution given the measurement $y$. This is achieved by setting $p_A$ and $p_B$ as follows:
\begin{equation}
p_A\left(X_0\right)=q_{clean}\left(X_0|X_{corrupt},y\right),
\label{eq:pA}
\end{equation}
\begin{equation}
p_B\left(X_1\right)=\delta\left(X_1-X_{corrupt}\right),
\label{eq:pB}
\end{equation}
where $q_{clean}(\cdot|X_{corrupt},y)$ represents the clean image distribution given specific $X_{corrupt}$ and $y$. 
\begin{theorem}
If $p_A$ and $p_B$ are defined according to equations (\ref{eq:pA}) and (\ref{eq:pB}), and $f$ is set to $0$, then the PDEs (\ref{eq:PDE}) and (\ref{eq:PDE_margin}) have the following analytical solutions:
\begin{equation}
\Psi\left(X_t,t\right)=\mathcal{N}\left(X_t|X_{corrupt},\overline{\sigma}_t^2I\right),
\label{eq:psi}
\end{equation}
\begin{equation}
\hat{\Psi}\left(X_t,t\right)=\int\hat{\Psi}_{X_0}\left(X_t,t\right)q_{clean}\left(X_0|X_{corrupt},y\right)dX_0,
\label{eq:psi_hat}
\end{equation}
where
\begin{equation}
\hat{\Psi}_{X_0}\left(X_t,t\right)=C_{X_0}\mathcal{N}\left(X_t|X_0,\sigma_t^2I\right),
\label{eq:psi_hat_x0}
\end{equation}
and
\begin{equation}
C_{X_0}=\left(\sqrt{2\pi}\sigma_{1}\right)^d\exp{\left(\frac{\left(X_0-X_{corrupt}\right)^\mathsf{T}\left(X_0-X_{corrupt}\right)}{2\sigma_1^2}\right)}.
\label{eq:C_x0}
\end{equation}
\label{theorem 1}
\end{theorem}
We set $f$ to $0$, and based on Theorem \ref{theorem 1}, $\Psi$ and $\hat{\Psi}$ can be analytically expressed as equations (\ref{eq:psi}) and (\ref{eq:psi_hat}). The gradient of $\log\Psi$ can be computed as:
\begin{equation}
\nabla\log\Psi=-\frac{1}{\overline{\sigma}_t^2}\left(X_t-X_{corrupt}\right).
\label{eq:gradient logPsi}
\end{equation}
By incorporating equation (\ref{eq:gradient logPsi}) into the forward SDE (\ref{eq:sde_forw}), we derive the forward process of MESB, which is same to that of I$^2$SB. 

For the reverse process of MESB, we need to parameterize $q_{clean}(X_0|X_{corrupt},y)$. This conditional probability is influenced not only by the distance between $X_0$ and the plane $AX=y$ according to equation (\ref{eq:inverse problem}), but also by the distance between $X_0$ and $Z\left(X_{corrupt},y\right)$, where $Z$ represents a function that can approximately map $X_{corrupt}$ and $y$ to their corresponding clean image. Thus, we assume that $q_{clean}(X_0|X_{corrupt},y)$ can be represented as:
\begin{equation}
q_{clean}\left(X_0|X_{corrupt},y\right)=k\mathcal{N}\left(X_0|Z\left(X_{corrupt},y\right),\Sigma_X\right)\mathcal{N}\left(y|AX_0,\sigma_y^2I\right).
\label{eq:q_clean}
\end{equation}
Here, $k$ is a constant independent of $X_0$ and $X_t$. The covariance matrices $\Sigma_X$ and $\sigma_y^2I$ reflect the confidence level in the approximate mapping $Z$ and the accuracy of the measurement $y$, respectively. Under this assumption, we can compute the gradient of $\log\hat{\Psi}$ as:
\begin{equation}
\nabla\log\hat{\Psi}=-\frac{1}{\sigma_t^2}\left(X_t-\hat{X}_{0,new}\right).
\label{eq:gradient logPsihat}
\end{equation}
In this equation, $\hat{X}_{0,new}$ is the solution of the linear equation:
\begin{equation}
M_t\hat{X}_{0,new}=\left(\left(X_t-\frac{\sigma_t^2}{\sigma_1^2}X_{corrupt}\right)+\sigma_t^2\Sigma_X^{-1}Z\left(X_{corrupt},y\right)+\frac{\sigma_t^2}{\sigma_y^2}A^\mathsf{T}y\right),
\label{eq:X_0new}
\end{equation}
where $M_t$ is defined as:
\begin{equation}
M_t=\left(\left(1-\frac{\sigma_t^2}{\sigma_1^2}\right)I+\sigma_t^2\Sigma_X^{-1}+\frac{\sigma_t^2}{\sigma_y^2}A^\mathsf{T}A\right).
\label{eq:Mt}
\end{equation}

By integrating equation (\ref{eq:gradient logPsihat}) into the backward SDE (\ref{eq:sde_back}), we derive the reverse process of MESB. Specifically, transitioning from $X_n$ to $X_{n-1}$ involves computing $\hat{X}_0$ using equation (\ref{eq:x0_hat}), and then solving the linear equation:
\begin{equation}
M_{t_n}\hat{X}_{0,new}=\left(\left(X_n-\frac{\sigma_n^2}{\sigma_N^2}X_{corrupt}\right)+\sigma_n^2\Sigma_X^{-1}\hat{X}_0+\frac{\sigma_n^2}{\sigma_y^2}A^\mathsf{T}y\right),
\label{eq:X_0new_discrete}
\end{equation}
where we substitute the approximate mapping $Z$ in equation (\ref{eq:X_0new}) with $\hat{X}_0$. Subsequently, $X_{n-1}$ is sampled from $p\left(X_{n-1}|\hat{X}_{0,new},X_n\right)$ according to equation (\ref{eq:p}). In practice, obtaining the exact solution of the linear equation (\ref{eq:X_0new_discrete}) is time-consuming; hence, we perform a $p$-th Conjugate Gradient (CG) update starting from $\hat{X}_0$. The generative process of MESB is summarized in Algorithm (\ref{alg1}).

\begin{algorithm}
    \renewcommand{\algorithmicrequire}{\textbf{Input:}}
    \renewcommand{\algorithmicensure}{\textbf{Initialize:}}
    \caption{Generative Process of MESB}
    \label{alg1}
    \begin{algorithmic}
		\REQUIRE $N$, $\left\{t_n\right\}$, $X_{corrupt}$, measurement $y$, system matrix A, trained network $\epsilon_{\theta^*}$
            \ENSURE $X_N=X_{corrupt}$
		\FOR{$n=N$ to $1$} 
		\STATE Predict $\hat{X}_0$ using $X_n$ and $\epsilon_{\theta^*}\left(X_n, t_n\right)$ according to equation (\ref{eq:x0_hat})
            \STATE Start from $\hat{X}_0$ and perform $p$-th CG update for linear equation (\ref{eq:X_0new_discrete}) to get $\hat{X}_{0,new}$	 
		\STATE Sample $X_{n-1}$ from $p\left(X_{n-1}|\hat{X}_{0,new}, X_n\right)$ according to equation (\ref{eq:p})
            \ENDFOR
        \RETURN $X_0$
    \end{algorithmic}
\end{algorithm}
\subsection{Understanding of Measurement Embedded Schrödinger Bridge}
\subsubsection{Simplification}
For a clearer understanding of MESB, we delve into equation (\ref{eq:X_0new_discrete}), which incorporates data consistency into the generative process. This equation can be viewed as an optimization problem:
\begin{equation}
\hat{X}_{0,new}=\arg\min_{X}\Vert X-\hat{X}_0\Vert _2^2+k_y\Vert AX-y\Vert_2^2+k_e\Vert X-X_{0,e}\Vert_2^2+\Vert T(X-\hat{X}_0)\Vert_2^2,
\label{eq:optimization}
\end{equation}
where 
\begin{equation}
X_{0,e} = \frac{\sigma_N^2}{\overline{\sigma}_n^2}X_n-\frac{\sigma_n^2}{\overline{\sigma}_n^2}X_{corrupt},
\label{eq:x0e}
\end{equation}
\begin{equation}
k_e=\frac{\overline{\sigma}_n^2\sigma_X^2}{\sigma_n^2\sigma_N^2},
\label{eq:ke}
\end{equation}
and
\begin{equation}
k_y=\frac{\sigma_X^2}{\sigma_y^2}.
\label{eq:ky}
\end{equation}
In these equations, $\sigma_X^2$ represents the largest eigenvalue of $\Sigma_X$, and $T$ denotes a transformation matrix associated with $\Sigma_X$ as described by the equation:
\begin{equation}
\Sigma_X^{-1}=\frac{1}{\sigma_x^2}\left(I+T^\mathsf{T}T\right).
\label{eq:T}
\end{equation}
According to equation (\ref{eq:optimization}), MESB considers four essential terms in the generative process. The first term $\Vert X-\hat{X}_0\Vert_2^2$ aims to keep $\hat{X}_{0,new}$ close to $\hat{X}_0$ and acts as a regularization term, particularly effective when dealing with noisy measurements. The second term $\Vert AX-y\Vert_2^2$ enforces hard data consistency within $\hat{X}_{0,new}$. The third term $\Vert X-X_{0,e}\Vert_2^2$ uses information from the extrapolation term $X_{0,e}$, which might capture additional details beyond the expected mean $\hat{X}_0$. Lastly, the fourth term $\Vert T(X-\hat{X}_0)\Vert_2^2$ ensures that $\hat{X}_{0,new}$ aligns with $\hat{X}_0$ under the transformation $T$, enabling the integration of prior knowledge about the clean image distribution into MESB through the design of the transformation matrix $T$.
\subsubsection{Relationship with CDDB and CDDB deep}
CDDB and CDDB-deep\cite{chung2024direct} are Schrödinger Bridge-based inverse problem solvers that use the same trained model as I$^2$SB and incorporate data consistency during sampling, similar to the techniques used in Decomposed Diffusion Sampling (DDS)\cite{chung2023fast} and Diffusion Posterior Sampling (DPS)\cite{chung2022diffusion}. Specifically, in each reverse step from  $X_n$ to $X_{n-1}$, CDDB updates the expected mean $\hat{X}_0$  by:
\begin{equation}
\hat{X}_{0,new}^{CDDB}=\hat{X}_0+\alpha A^\mathsf{T}\left(y-A\hat{X_0}\right),
\label{eq:CDDB}
\end{equation}
and CDDB-deep updates it by: 
\begin{equation}
\hat{X}_{0,new}^{deep}=\hat{X}_0-\alpha\nabla_{X_n}\Vert A\hat{X}_0(X_n)-y\Vert_2^2,
\label{eq:CDDB deep}
\end{equation}
where the step lengths 
$\alpha$ are treated as hyperparameters. These updated means, $\hat{X}_{0,new}^{CDDB}$ and $\hat{X}_{0,new}^{deep}$, are then used for sampling $X_{n-1}$ according to equation (\ref{eq:p}). Here, we establish the connections between our proposed MESB and CDDB as well as CDDB-deep.
\begin{theorem}
If the system matrix $A$ satisfies the following two conditions: firstly, there exists a positive number $\alpha_0$ such that $\sqrt{\alpha_0}A$ is a partially isometric matrix\cite{garcia2019partially}, and secondly, $A$ is row full rank, then the CDDB update for $\hat{X}_0$ in equation (\ref{eq:CDDB}) is equivalent to solving the optimization problem:
\begin{equation}
\hat{X}_{0,new}^{eq}=\arg\min_X\Vert X-\hat{X}_0\Vert_2^2+k\Vert AX-y\Vert_2^2.
\label{eq: CDDB_equivalent}
\end{equation}
Specifically, when $\alpha=\frac{\alpha_0k}{\alpha_0+k}$, $\hat{X}_{0,new}^{CDDB}=\hat{X}_{0,new}^{eq}$.
\label{theorem 2}
\end{theorem}
In applications such as pool super-resolution, inpainting, and MRI acceleration, where the system matrix $A$ strictly satisfies the conditions described in Theorem \ref{theorem 2}, CDDB can be considered as a special case of our proposed MESB. This is because the update for $\hat{X}_0$ in MESB becomes equivalent to the CDDB update when setting $k_e$ and $T$ in equation (\ref{eq:optimization}) to $0$.  In other applications, such as deblurring and sparse-view CT reconstruction, where the system matrix $A$ does not satisfy these conditions, a single-step gradient update cannot fully incorporate all the information from the measurements into  $\hat{X}_0$.  In such cases, our proposed MESB, which uses hard data consistency along with regularization terms, can achieve better performance than CDDB.
\begin{theorem}
If $p_A$ and $p_B$ are defined according to equations (\ref{eq:pA}) and (\ref{eq:pB}),  $f$ is set to $0$, $\Psi$ and $\hat{\Psi}$ are expressed in equations (\ref{eq:psi}) and (\ref{eq:psi_hat}), and $X_t$ is sampled from equation (\ref{eq:mu_t}), assuming that given $X_t$ and $X_{corrupt}$, the measurement $y$ follows a gaussian distribution centered at $A\hat{X}_0\left(X_t\right)$:
\begin{equation}
q\left(y|X_t,X_{corrupt}\right)=\mathcal{N}\left(y|A\hat{X}_0\left(X_t\right),\sigma^2I\right),
\label{eq:deep_y}
\end{equation}
where $\hat{X}_0\left(X_t\right)$ is the expected mean, 
then the gradient of $\log\hat{\Psi}$ can be expressed as:
\begin{equation}
\nabla\log\hat{\Psi}=-\frac{1}{\sigma_t^2}\left(X_t-\hat{X}_{0,new}^{deep}\right),
\label{eq:gradient_logpsihat_deep}
\end{equation}
where $\hat{X}_{0,new}^{deep}$ is expressed in equation (\ref{eq:CDDB deep}), and the step length $\alpha$  equals to $\frac{\sigma_t^2}{2\sigma^2}$.
\label{theorem 3}
\end{theorem}
As indicated by Theorem \ref{theorem 3}, CDDB deep can also be considered as establishing Schrödinger Bridges between $p_A$ and $p_B$ as defined in the equations (\ref{eq:pA}) and (\ref{eq:pB}), but with a different assumption to decouple $\hat{\Psi}$ and $q_{clean}\left(X_0|X_{corrupt},y\right)$ in the equation (\ref{eq:psi_hat}). The incorporation of the U-net Jacobian in CDDB deep makes it suitable for tasks like inpainting, where a global impact on all pixels is desired. However, this incorporation also makes it more time- and memory-consuming than our proposed MESB.

\subsection{Implementation Details}
\begin{figure}[t]
  \centering
  \includegraphics[width=\textwidth]{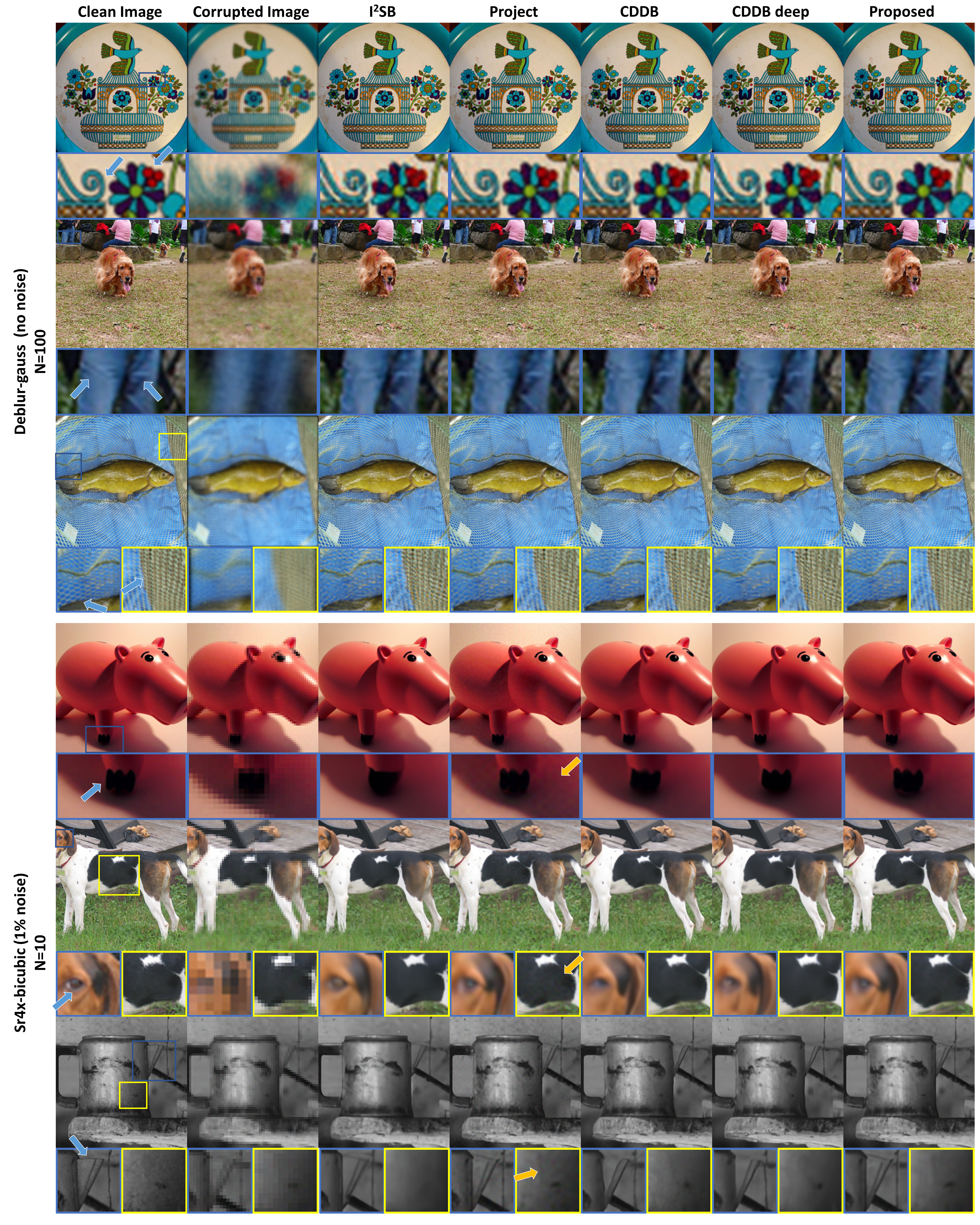}
  \caption{Visualization results for the deblur-gauss (no noise) task and the sr4x-bicubic (1\% noise) task. The details within the blue and yellow boxes are zoomed in for enhanced visual clarity.}
  \label{fig1}
\end{figure}

\begin{figure}
  \centering
  \includegraphics[width=\textwidth]{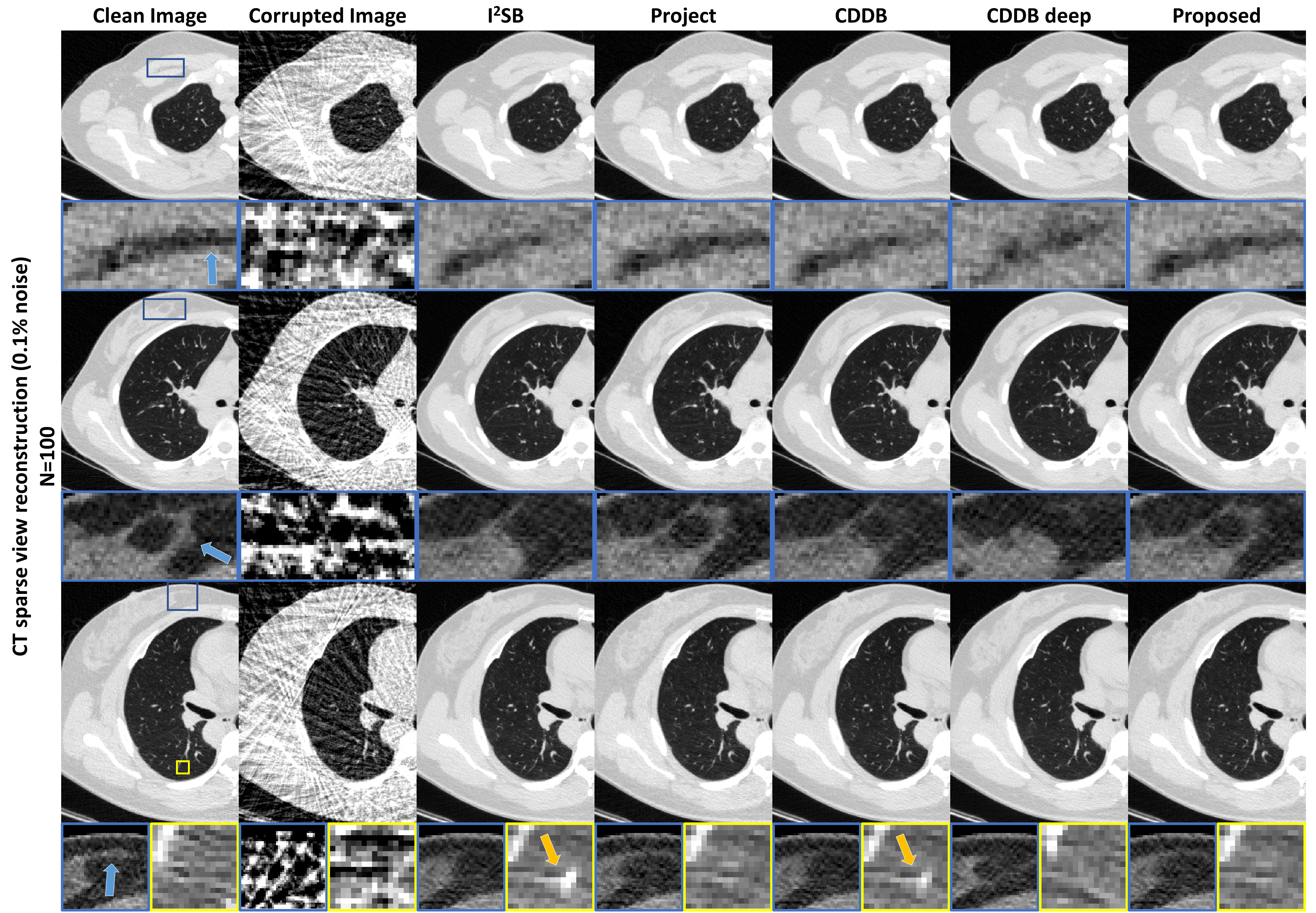}
  \caption{Visualization results for the CT sparse view reconstruction (0.1\% noise) task. The details within the blue and yellow boxes are zoomed in for enhanced visual clarity. The display window for the entire images is set to [-1000HU, 200HU], for the zoomed regions outside the lungs is set to [-160HU, 200HU], and for the zoomed regions inside the lungs is set to [-1000HU, -550HU].}
  \label{fig2}
\end{figure}

\begin{table}[t]
\begin{center}
\caption{
Quantitative results of tested methods for the deblur-gauss (no noise) task. \textbf{Bold}: best, \underline{under}: second best.}\label{tab1}
\setlength{\tabcolsep}{2.4mm}{
\begin{tabular}{c|cccc|cccc}
\hline
 &\multicolumn{4}{c|}{LPIPS (Corrupt: 0.4291)}&\multicolumn{4}{c}{SSIM (Corrupt: 0.5995)} \\
$N$       &10&20         &50 &100          &10&20  &50         & 100   \\
\hline
I$^2$SB      & 0.0620           & 0.0617        &0.0622 & 0.0625           & 0.9490        &0.9475 &0.9463 &0.9458    \\
Project       & \underline{0.0454}          & \underline{0.0437}          & \underline{0.0424}   & \underline{0.0417}           & \underline{0.9689}         & \underline{0.9702} &\underline{0.9714}&\underline{0.9722}   \\
CDDB    & 0.0538          &0.0525          & 0.0523 &0.0523            & 0.9593          & 0.9595          &  0.9594 & 0.9594    \\
CDDB deep & 0.0611&0.0601&0.0605&0.0629&0.9498&0.9491&0.9486&0.9477\\
Proposed & \textbf{0.0440 }&\textbf{0.0414}&\textbf{0.0398}&\textbf{0.0390}&\textbf{0.9716}&\textbf{0.9737}&\textbf{0.9748}&\textbf{0.9754}\\
\hline

\end{tabular}}
\end{center}
\end{table}

\begin{table}[t]
\begin{center}
\caption{
Quantitative results of tested methods for the sr4x-bicubic (1\% noise) task. \textbf{Bold}: best, \underline{under}: second best.}\label{tab2}
\setlength{\tabcolsep}{2.4mm}{
\begin{tabular}{c|cccc|cccc}
\hline
 &\multicolumn{4}{c|}{LPIPS (Corrupt: 0.4693)}&\multicolumn{4}{c}{SSIM (Corrupt: 0.6633)} \\
$N$       &10&20         &50 &100          &10&20  &50         & 100   \\
\hline
I$^2$SB      & 0.2772           & 0.2698        &0.2633 & 0.2611          & 0.7490        &0.7320 &0.7111 &0.6987    \\
Project       & 0.2433          & 0.2348          & 0.2302   & 0.2314          & 0.7789        & 0.7643&0.7432&0.7289   \\
CDDB    & \underline{0.2409}          &\underline{0.2294}          & 0.2197 &\underline{0.2171}            & \underline{0.7793}          & 0.7674          &  0.7505 & 0.7395    \\
CDDB deep & 0.2529&0.2329&\textbf{0.2105}&\textbf{0.2082}&0.7755&\textbf{0.7712}&\textbf{0.7613}&\textbf{0.7452}\\
Proposed & \textbf{0.2361}&\textbf{0.2258}&\underline{0.2186}&0.2182&\textbf{0.7829}&\underline{0.7703}&\underline{0.7521}&\underline{0.7400}\\
\hline

\end{tabular}}
\end{center}
\end{table}

\begin{table}[t]
\begin{center}
\caption{
Quantitative results of tested methods for CT sparse view reconstruction (0.1\% noise ) task. \textbf{Bold}: best, \underline{under}: second best}\label{tab3}
\setlength{\tabcolsep}{2.4mm}{
\begin{tabular}{c|cccc|cccc}
\hline
 &\multicolumn{4}{c|}{LPIPS (Corrupt: 0.5034)}&\multicolumn{4}{c}{SSIM (Corrupt: 0.3193)} \\
$N$       &10&20         &50 &100          &10&20  &50         & 100   \\
\hline
I$^2$SB      & 0.2236          & 0.2082       &0.1910 & 0.1838          & 0.9192        &0.9141 &0.9055 &0.8981    \\
Project       & \underline{0.2199}          & 0.2087          & 0.1975   & 0.1938          & \underline{0.9208}        & \underline{0.9171}&\underline{0.9104}&\underline{0.9041}   \\
CDDB    & 0.2219          &\underline{0.2065}         & \underline{0.1893} &0.1815            & 0.9203          & 0.9160          &  0.9075 & 0.9020    \\
CDDB deep & 0.2226&0.2072&0.1897&\underline{0.1810}&0.9206&0.9164&0.9075&0.9006\\
Proposed & \textbf{0.2144}&\textbf{0.2009}&\textbf{0.1865}&\textbf{0.1801}&\textbf{0.9238}&\textbf{0.9201}&\textbf{0.9134}&\textbf{0.9074}\\
\hline

\end{tabular}}
\end{center}
\end{table}
We validated our proposed method on both natural and medical image tasks. For natural image tasks, we used the pretrained model of I$^2$SB\cite{liu20232} and evaluated our proposed MESB with 5,000 images randomly selected from the validation dataset of ImageNet 256x256\cite{deng2009imagenet}. We tested MESB on the following degradation tasks: gaussian deblurring with no added noise and 4x super-resolution with bicubic interpolation and $1\%$ gaussian noise.

For the medical image tasks, we validated our proposed MESB with CT sparse view reconstruction. We used the RPLHR-CT-tiny dataset\cite{yu2022rplhr}, consisting of anonymized chest CT volumes. The original CT images served as ground truth, and the corresponding corrupted images were obtained using the FBP algorithm with projections from 60 distinct views in a fan beam geometry. We used 40 cases (11,090 slices) with 0.01\% Gaussian noise added in the projections for training and 5 cases (1,425 slices) with 0.1\% Gaussian noise added in the projections for testing. The neural network $\epsilon_{\theta}\left(X_n,t_n\right)$ we used is a 2D residual U-Net with the same architecture used in DDPM\cite{ho2020denoising}. We concatenated positional encoded $X_{corrupt}$ with $X_n$ along the channel dimension to serve as additional conditions for the network. During training, we used 1000 diffusion time steps with quadratic discretization, and adopted a symmetric scheduling\cite{chen2021likelihood,de2021diffusion} of $\beta_t$ with a maximum value of $0.15$ at $t=0.5$. The model was trained on randomly cropped patches of size 128 $\times$ 128 and tested on the entire 512 $\times$ 512 images. A batch size of 64 was employed during training, using the Adam algorithm with a learning rate of $8\times10^{-5}$ for 200,000 iterations. The number of generative steps $N$ was set to 10, 20, 50, and 100, and the time of CG iterations $p$ for each reverse step was set to 5.

We specify the corrupted image $X_{corrupt}$ and the measurement $y$ for each task. In the deblur-gauss (no noise) task, $y$ and $X_{corrupt}$ are both blurred images. In the sr4x-bicubic (1\% noise) task, $y$ is the downsampled image, and $X_{corrupt}$ is the reconstructed image from $y$ using nearest neighbor interpolation.  In the CT sparse view (0.1\% noise) task, $y$ represents the projections and $X_{corrupt}$ is the reconstructed image from $y$ using FBP algorithm.

The hyperparameters $k_e$ and $k_y$ and the transformation matrix $T$ are set as follows. In the deblur-gauss (no noise) task, $k_y$ is set to positive infinity, $k_e$ is set to $20\frac{\sigma_n^2\overline{\sigma}_n^2}{\sigma_N^4}$, and $T$ is set to 0. In the sr4x-bicubic (1\% noise) task, $k_y$ is set to 32, $k_e$ is set to 0, and $T$ is set to 0. In the CT sparse view reconstruction (0.1\% noise) task, the Frobenius norm of $A$ is 2051.5, $k_y$ is set to 0.01, $k_e$ is set to 0, and $T^\mathsf{T}T$ is set to $-0.5\triangle$, where $\triangle$ denotes a 2 dimensional discrete Laplacian operator. 
\section{Results}
We conducted a comparative analysis between I$^2$SB, CDDB, CDDB deep, and our proposed MESB. Additionally, we included another comparison method degenerated from MESB by setting $k_y$ to positive infinity, and $k_e$ and $T$ to $0$. We term this method "Project" since, in each reverse step, it projects $\hat{X}_0$ onto the plane $AX=y$.  The number of CG iterations for Project in each reverse step is set to 5. For a fair comparison, all tested methods use the same trained models, and the hyperparameters during testing are optimized to the best of our ability. The hyperparameter settings for CDDB and CDDB deep are detailed in Appendix A.2.2. Representative results for natural images are visualized in Figure \ref{fig1}, and medical images in Figure \ref{fig2}. For quantitative analysis, we calculated learned perceptual image patch similarities (LPIPS)\cite{zhang2018unreasonable} to evaluate the texture restoration ability of the tested methods and structural similarity index measures (SSIMs) to assess the fidelity of the tested methods to the ground truth. These quantitative results are detailed in Tables \ref{tab1}, \ref{tab2}, and \ref{tab3}. See Appendix A.3 for additional results, including ablation studies and statistical significance tests. 

In the deblur-gauss (no noise) task, our proposed MESB consistently outperformed all other methods across all generative steps $N$. MESB significantly outperforms I$^2$SB, CDDB, and CDDB deep, with improvements of 20\% to 30\% in LPIPS and increases of 0.015 to 0.025 in SSIMs. This is further supported by the visualization results in Figure \ref{fig1}, where MESB shows superior restoration details in areas like cherries, pant pleats, and fishing nets. Compared to the second-place method, Project, our proposed MESB achieved about a 3\% improvement in LPIPS and a 0.003 increase in SSIMs. 

In the sr4x-bicubic (1\% noise) task, when $N$ equals to 10, our proposed MESB achieves the best results in terms of both LPIPS and SSIMs. As visualized in Figure \ref{fig1}, MESB demonstrates superior restoration details in areas like pig's knuckles, dog's eyes, and cup mouths. Since the system matrix in this task nearly satisfies the conditions described in Theorem \ref{theorem 2}, the performance difference between MESB and CDDB decreases as the generative steps N increase. For $N$ equals to 50 and 100, CDDB deep achieves the best results, but the computation of the U-net Jacobian makes it approximately three times more computationally expensive than other methods (see Appendix A.2.1 for computation time). Project does not perform well in this task due to its sensitivity to noise. As highlighted by the yellow arrows in Figure \ref{fig1}, Project introduces unrealistic artifacts. Compared to Project, MESB achieves a 5\% to 10\% improvement in LPIPS and a 0.005 to 0.01 increase in SSIMs.

In the CT sparse view reconstruction (0.1\% noise) task, our proposed MESB consistently outperformed all other methods across all generative steps $N$. Compared with CDDB and CDDB deep, MESB achieved a 0.5\% to 4\% improvement in LPIPS and a 0.003 to 0.007 increase in SSIMs. As shown in Figure \ref{fig2}, MESB demonstrates superior detail restoration in mammary glands and corrects the inaccurately generated pulmonary vein seen in I$^2$SB. Project achieved second place in terms of SSIM, but its LPIPS were worse than those of I$^2$SB due to overfitting to noise in the measurement. Compared with Project, MESB achieved a 2\% to 7\% improvement in LPIPS and a 0.003 increase in SSIMs.

Overall, while the performances of CDDB and CDDB deep are significantly influenced by the system matrix, and Project is sensitive to noise in measurements, our proposed MESB consistently achieves good performance across all three tasks.

\section{Conclusion and Discussion}
In conclusion, we propose MESB, a novel model that establishes Schrödinger Bridges between the distribution of corrupted images and the distribution of clean images given measurements. Based on optimal transport theory, we embed measurements into the marginal condition of the Schrödinger Bridge, deriving both the forward and backward processes. We also provide an explanation of MESB and its connections with CDDB and CDDB deep. Our proposed MESB demonstrates robustness to noise and outperforms existing Schrödinger Bridge-based inverse problem solvers in both natural and medical image tasks.

MESB shows promise for further refinement in the design of the transformation matrix $T$. Currently, $T$ is set to 0 for natural images and as a gradient operator for CT images. A more dedicated design of $T$ may incorporate additional prior knowledge and further enhance the performance of MESB. Additionally, while measurements are not necessary during training, paired clean and corrupted images are required. In future work, we will explore extending MESB to unpaired data and broadening the application scenarios for MESB.

%\bibliographystyle{IEEEtran}
%\bibliography{mybibliography}

%%%%%%%%%%%%%%%%%%%%%%%%%%%%%%%%%%%%%%%%%%%%%%%%%%%%%%%%%%%%

\appendix
\section{Appendix}

\subsection{Proof}
\subsubsection{Proof of Theorem \ref{theorem 1}}
To prove Theorem \ref{theorem 1}, we first verify that the $\Psi$ expressed in equation (\ref{eq:psi}) satisfies:
\begin{equation}
\frac{\partial\Psi}{\partial t}=-\frac{1}{2}\beta\triangle\Psi.
\end{equation}
This is because:
\begin{equation}
\begin{aligned}
\frac{\partial\Psi}{\partial t} &= \frac{\partial{\Psi}}{\partial\overline{\sigma}_t^2}\frac{\partial\overline{\sigma}_t^2}{\partial t},\\
&=-\frac{1}{2}\beta\Psi\left(\frac{\left(X_t-X_{corrupt}\right)^\mathsf{T}\left(X_t-X_{corrupt}\right)}{\overline{\sigma}_t^4}-\frac{d}{\sigma_t^2}\right),\\
&=-\frac{1}{2}\beta\triangle\Psi.
\end{aligned}
\end{equation}
In a similar way, $\hat\Psi_{X_0}\left(X_t,t\right)$ satisfies:
\begin{equation}
\frac{\partial\hat{\Psi}_{X_0}}{\partial t}=\frac{1}{2}\beta\triangle\hat{\Psi}_{X_0}.
\end{equation}
Hence, the $\hat{\Psi}$ expressed in equation (\ref{eq:psi_hat}) satisfies:
\begin{equation}
\begin{aligned}
\frac{\partial\hat{\Psi}}{\partial t}&=\int\frac{\partial\hat{\Psi}_{X_0}}{\partial t}q_{clean}\left(X_0|X_{corrupt},y\right)dX_0, \\
&=\frac{1}{2}\beta\triangle\hat{\Psi}.
\end{aligned}
\end{equation}
Therefore, $\Psi$ and $\hat{\Psi}$ satisfy the PDEs (\ref{eq:PDE}). For the margin conditions (\ref{eq:PDE_margin}),
\begin{equation}
\Psi\left(X_0,0\right)=\frac{1}{C_{X_0}},
\end{equation}
where $C_{X_0}$ is expressed in equation (\ref{eq:C_x0}), and
\begin{equation}
\begin{aligned}
\hat\Psi\left(X_0,0\right)&=\int C_X\delta\left(X_0-X\right)q_{clean}\left(X|X_{corrupt},y\right)dX, \\
&=C_{X_0}q_{clean}\left(X_0|X_{corrupt},y\right),
\end{aligned}
\end{equation}
then 
\begin{equation}
\Psi\left(X_0,0\right)\hat\Psi\left(X_0,0\right)=p_A,
\end{equation}
where $p_A$ is defined as equation (\ref{eq:pA}). Also
\begin{equation}
\Psi\left(X_1,1\right)=\delta\left(X_1-X_{corrupt}\right),
\end{equation}
and because  $\hat{\Psi}_{X_0}\left(X_1=X_{corrupt},1\right)$ equals to 1, we have:
\begin{equation}
\begin{aligned}
\Psi\left(X_1,1\right)\hat{\Psi}\left(X_1,1\right)&=\delta\left(X_1-X_{corrupt}\right)\int\hat{\Psi}_{X_0}\left(X_1,1\right)q_{clean}\left(X_0|X_{corrupt},y\right)dX_0, \\
&=\delta\left(X_1-X_{corrupt}\right), \\
&=p_B.
\end{aligned}
\end{equation}
That completes the proof.
\begin{table}[t]
\begin{center}
\caption{
The sampling time of tested methods on different tasks. }\label{tab5}
\setlength{\tabcolsep}{4mm}{
\begin{tabular}{c|ccccc}
\hline
sec./(iter $\cdot$ batch size)   &I$^2$SB&Project  &CDDB &CDDB deep&Proposed            \\
\hline
Deblur-gauss      & 0.031          & 0.032       &0.031 & 0.096  &  0.032         \\
Sr4x-bicubic      &0.031           & 0.032          &0.031   &0.096    & 0.032      \\
CT sparse view    & 0.043         &0.060         & 0.050 &0.184   &  0.060         \\
\hline

\end{tabular}}
\end{center}
\end{table}
\subsubsection{Proof of equation (\ref{eq:gradient logPsihat})}
This equation can be proved by substituting equation (\ref{eq:q_clean}) into equation (\ref{eq:psi_hat}), calculating the integration regarding to $X_0$ and finally calculating the log gradient regarding to $X_t$.
\subsubsection{Proof of Theorem \ref{theorem 2}}
According to the definition of the partially isometric matrix, we have:
\begin{equation}
A=\alpha_0AA^{\mathsf{T}}A.
\label{eq:A}
\end{equation}
Also because $A$ is row full rank, for any $y$, there always exists an $X_0$ such that $y=AX_0$.
We use $f\left(X\right)$ to represent:
\begin{equation}
f\left(X\right)=\Vert X-\hat{X}_0\Vert_2^2+k\Vert A(X-X_0)\Vert_2^2,
\end{equation}
then
\begin{equation}
\frac{df}{dX}=2\left(\left(X-\hat{X}_0\right)+kA^\mathsf{T}A\left(X-X_0\right)\right).
\label{eq:dfdx}
\end{equation}
According to equation (\ref{eq:CDDB}), we have
\begin{equation}
\hat{X}_{0,new}^{CDDB}=\hat{X}_0+\alpha A^\mathsf{T}A\left(X_0-\hat{X_0}\right).
\label{eq:CDDB_44}
\end{equation}
Using equation \ref{eq:A} and substituting $X$ in equation (\ref{eq:dfdx}) with $\hat{X}_{0,new}^{CDDB}$, we figure out that if $\alpha=\frac{\alpha_0k}{\alpha_0+k}$, $\frac{df}{dX}$ equals to 0 when $X$ equals to $\hat{X}_{0,new}^{CDDB}$. Therefore, $\hat{X}_{0,new}^{CDDB}$ is the optimal point for minimizing $f\left(X\right)$. That completes the proof.
\subsubsection{Proof of Theorem \ref{theorem 3}}
To prove theorem \ref{theorem 3}, we have
\begin{equation}
\nabla\log\hat{\Psi}=\nabla\log\left(\hat{\Psi}\Psi\right)-\nabla\log\Psi.
\end{equation}
Noting that
\begin{equation}
q\left(X_t|X_0,X_{corrupt}\right)=k_{X_t}\Psi\hat{\Psi}_{X_0},
\end{equation}
where $q$ is defined in equation (\ref{eq:mu_t}), $X_1$ is substituted by $X_{corrupt}$, and $k_{X_t}$ denotes a constant independent of $X_0$, we have:
\begin{equation}
\begin{aligned}
\nabla\log\hat{\Psi}=\nabla\log\left(\frac{1}{k_{X_t}}\int q\left(X_t|X_0,X_{corrupt}\right)q_{clean}\left(X_0|X_{corrupt},y\right)dX_0\right)-\nabla\log\Psi.
\end{aligned}
\end{equation}
Given $X_0$ and $X_{corrupt}$, $X_t$ is independent of $y$,
therefore
\begin{equation}
\begin{aligned}
\nabla\log\hat{\Psi}&=-\nabla\log k_{X_t}+\nabla\log q\left(X_t|X_{corrupt},y\right)-\nabla\log\Psi, \\
&=-\nabla\log k_{X_t}+\nabla\log q\left(X_t|X_{corrupt}\right)-\nabla\log\Psi+\nabla\log q\left(y|X_t,X_{corrupt}\right),\\
&=\nabla\log\left(\int\hat{\Psi}_{X_0}q_{clean}\left(X_0|X_{corrupt}\right)dX_0\right)+\nabla\log q\left(y|X_t,X_{corrupt}\right).
\end{aligned}
\end{equation}
The first term independent of measurement $y$ is learned by score matching and the second term is computed using the assumption (\ref{eq:deep_y}). Therefore,
\begin{equation}
\nabla\log\hat{\Psi}=-\frac{1}{\sigma_t^2}\left(X_t-\hat{X}_0\left(X_t\right)\right)-\frac{1}{2\sigma^2}\nabla_{X_t}\Vert A\hat{X}_0(X_t)-y\Vert_2^2,
\end{equation}
where the expected mean $\hat{X}_0$ is defined as:
\begin{equation}
\hat{X}_0=\int X_0q\left(X_0|X_t,X_{corrupt}\right)dX_0.
\end{equation}
That completes the proof.

\subsection{Experimental Details}
\subsubsection{Running Statistics}
All experiments were run using a single A100-SXM4-40GB GPU. The sampling time of tested methods on different tasks is detailed in Table \ref{tab5}.

\subsubsection{Hyperparameters for Comparison Methods}
In the deblur-gauss (no noise) task, the step length $\alpha$ is set to 10 for CDDB and 0.01 for CDDB deep. In the sr4x-bicubic (1\% noise) task, $\alpha$ is set to 10 for CDDB and 4 for CDDB deep. For the CT sparse view reconstruction (0.1\% noise) task, $\alpha$ is set to 0.001 for CDDB and $\frac{0.05}{\Vert A\hat{X}_0-y\Vert_2}$ for CDDB deep.
\subsection{Additional Results}
\subsubsection{Ablation Studies}
We conducted ablation studies on the hyperparameters $k_y$ and $k_e$ of our proposed MESB. First, we examined the impact of $k_y$ in CT sparse view reconstruction (0.1\% noise) task. We randomly selected 32 slices from the test dataset, fixing $k_e$ to 0 and $T$ to $-0.5\triangle$. We varied $k_y$ from 0.0025 to 0.05, with quantitative results detailed in Table \ref{tab6}. As $k_y$ increases, LPIPS and SSIMs initially become better and then go worse for all generative steps $N$, indicating that an optimal $k_y$ enhances the performance of MESB. 

Next, we studied the impact of $k_e$ in deblur-gauss (no noise) task. We randomly selected 180 natural images from the validation dataset of  Imagenet 256x256, and fixed $k_y$ to positive infinity and $T$ to $0$. We set $k_e$ to $k_E\frac{\sigma_n^2\overline{\sigma}_n^2}{\sigma_N^4}$, with $k_E$ ranging from 0 to 100, and quantitative results detailed in Table \ref{tab7}. As $k_E$ increases, LPIPS and SSIMs initially become better and then go worse for all generative steps $N$, indicating that an optimal $k_E$ enhances the performance of MESB. 
\begin{table}[t]
\begin{center}
\caption{
Ablation study for $k_y$ in the CT sparse view reconstruction (0.1\% noise) task. \textbf{Bold}: Best.}\label{tab6}
\setlength{\tabcolsep}{2.4mm}{
\begin{tabular}{l|cccc|cccc}
\hline
 &\multicolumn{4}{c|}{LPIPS}&\multicolumn{4}{c}{SSIM} \\
$k_y$\verb|\|$N$      &10&20         &50 &100          &10&20  &50         & 100   \\
\hline
0.0025      & 0.2213          & 0.2063       &0.1902 & 0.1836          & 0.9143       &0.9096 &0.9014 &0.8947    \\
0.005       & 0.2176          & 0.2037          & \textbf{0.1889}   & \textbf{0.1826}         & 0.9151       & 0.9108&0.9029&0.8965  \\
0.01   & \textbf{0.2157}          &\textbf{0.2028}         & 0.1892 &0.1832            & \textbf{0.9154}          & \textbf{0.9112}          &  \textbf{0.9037} & \textbf{0.8973}    \\
0.025 &0.2165&0.2042&0.1916&0.1863&0.9147&0.9107&0.9034&0.8971 \\
0.05 & 0.2180&0.2061&0.1941&0.1892&0.9139&0.9100&0.9027&0.8963\\
\hline
\end{tabular}}
\end{center}
\end{table}

\begin{table}[t]
\begin{center}
\caption{
Ablation study for $k_E$ in the deblur gauss (no noise) task. \textbf{Bold}: Best.}\label{tab7}
\setlength{\tabcolsep}{2.4mm}{
\begin{tabular}{l|cccc|cccc}
\hline
 &\multicolumn{4}{c|}{LPIPS}&\multicolumn{4}{c}{SSIM} \\
$k_E$\verb|\|$N$      &10&20         &50 &100          &10&20  &50         & 100   \\
\hline
0      & 0.0451          & 0.0434       &0.0421 & 0.0413          & 0.9701       &0.9713 &0.9724 &0.9732    \\
10      & \textbf{0.0433}          & \textbf{0.0409}          & \textbf{0.0393}   & \textbf{0.0386}         & 0.9725       & 0.9741&0.9753&0.9758  \\
20   & 0.0435          & 0.0410          & 0.0394   & 0.0388         & 0.9727       & \textbf{0.9746}&\textbf{0.9757}&\textbf{0.9760}      \\
50 &0.0443&0.0423&0.0406&0.0399&\textbf{0.9729}&0.9744&0.9755&0.9760 \\
100 & 0.0455&0.0437&0.0423&0.0417&0.9727&0.9741&0.9751&0.9755\\
\hline
\end{tabular}}
\end{center}
\end{table}
\subsubsection{Statistical Significance Tests}
We performed dependent t-tests to compare the LPIPS and SSIMs of our proposed MESB with those of the comparison methods across different images. The p-values were all less than 0.04, indicating that the differences in mean values shown in Tables \ref{tab1}, \ref{tab2}, and \ref{tab3} are statistically significant.

\end{document}